\documentclass[journal=jacsat,manuscript=article]{achemso}

\setkeys{acs}{articletitle=true,etalmode=truncate,maxauthors=5}
\usepackage[version=3]{mhchem} 
\usepackage{caption}
\usepackage{float}
\usepackage{geometry}
\usepackage{xkeyval}
\usepackage{setspace}
\usepackage{tabularx}
\usepackage{graphicx}
\usepackage{dcolumn}
\usepackage{bm}
\usepackage{soul}
\usepackage{color}
\usepackage{xcolor}
\usepackage{amsmath}
\usepackage{amssymb}
\usepackage{amsmath}
\usepackage{natmove}
\usepackage{multirow}
\usepackage{array}
\usepackage{booktabs}
\usepackage{rotating}
\usepackage[normalem]{ulem}
\usepackage[colorlinks=true,linkcolor=blue,citecolor=blue,urlcolor=blue]{hyperref}
\usepackage{physics}
\usepackage{mathrsfs}
\usepackage[capitalise,noabbrev]{cleveref}
\usepackage{comment}
\usepackage{pdfpages}
\usepackage{hyperref}

\usepackage{subcaption}
\usepackage{caption}
\author{Michael Alejandro Hernandez Bertran}
\affiliation{Dipartimento di Scienze Fisiche, Informatiche, Matematiche (FIM), Universit\`a di Modena e Reggio Emilia, 41125 Modena, Italy}
\alsoaffiliation{Nanoscience Institute, National Research Council (CNR-NANO), 41125 Modena, Italy}
\author{Davide Tisi}
\affiliation{Laboratory of Computational Science and Modeling, Institut des Mat\'eriaux, \'Ecole Polytechnique F\'ed\'erale de Lausanne, 1015 Lausanne, Switzerland}
\author{Federico Grasselli}
\affiliation{Dipartimento di Scienze Fisiche, Informatiche, Matematiche (FIM), Universit\`a di Modena e Reggio Emilia, 41125 Modena, Italy}
\alsoaffiliation{Nanoscience Institute, National Research Council (CNR-NANO), 41125 Modena, Italy}
\author{Michele Ceriotti}
\affiliation{Laboratory of Computational Science and Modeling, Institut des Mat\'eriaux, \'Ecole Polytechnique F\'ed\'erale de Lausanne, 1015 Lausanne, Switzerland}
\author{Elisa Molinari}
\affiliation{Dipartimento di Scienze Fisiche, Informatiche, Matematiche (FIM), Universit\`a di Modena e Reggio Emilia, 41125 Modena, Italy}
\alsoaffiliation{Nanoscience Institute, National Research Council (CNR-NANO), 41125 Modena, Italy}
\author{Deborah Prezzi}
\affiliation{Nanoscience Institute, National Research Council (CNR-NANO), 41125 Modena, Italy}
\email{deborah.prezzi@cnr.it}

\title[]{Tracking the Lithiation State of Li$_x$Si from Machine-Learned XPS Binding Energies}

\abbreviations{IR,NMR,UV}

\begin{document}

\begin{tocentry}

\end{tocentry}

\begin{abstract}
X-ray Photoelectron Spectroscopy (XPS) is a powerful technique to probe chemical states and interfacial processes in battery materials, but a quantitative interpretation is often hindered by the complex, heterogeneous microstructures that form during operation and dominate electrochemical cycling.
Silicon based anodes represent a paradigmatic example in Li batteries, as (de)lithiation proceeds through the formation of strongly disordered Li$_x$Si phases and crystal-amorphous transformations that are hard to characterize.  
Here, we introduce a computational framework that combines machine-learning (ML) prediction of core-level binding energies to large-scale atomistic simulations -- Grand Canonical Monte Carlo (GCMC) complemented with molecular dynamics (MD), driven by a ML potential --  for a systematic sampling of lithiation states and local atomic environments. This approach yields stoichiometry maps that match the characteristic experimental trends observed in \textit{operando} and \textit{ex situ} XPS measurements, including the distinctive Si $2p$ spectroscopic signatures associated with the crystal-to-amorphous disordering driving early delithiation.
\end{abstract}

\section{Introduction}

Core-level spectroscopies are established tools/methods to assess the electronic structure and chemical environment of materials. Among these techniques, X-ray Photoelectron Spectroscopy (XPS) stands out as one of the most widespread, allowing both qualitative and quantitative analysis with information depths from a few to tens of nm depending on the radiation source\cite{weiland2013,woicik2016}. 
In battery research, XPS is routinely employed to characterize  electrode surfaces, providing insights into the oxidation state of redox-active species and the composition of the solid electrolyte interphase (SEI)\cite{fehse2021}. 
Recent advances in electrochemical cell design\cite{axnada2015,zhu2021} and the development of solid electrolytes\cite{manthiram2017} have enabled XPS measurements \textit{in situ}\cite{wenzel2015,favaro2019} and \textit{operando}\cite{kiuchi2020,mirolo2021,endo2022}, which can complement  standard \textit{ex-situ} studies~\cite{philippe2012,sachs2015,assat2017} by capturing interfacial and redox processes under realistic electrochemical conditions. During cycling, however, the probed materials can become highly heterogeneous and structurally complex, sampling disordered and metastable states for which quantitative spectral references are limited, making it difficult to translate measured binding-energy shifts into composition- and structure-resolved information. Extrinsic effects, such as evolving SEI contributions, local electrostatic potentials and differential charging, can further complicate energy referencing and peak assignment~\cite{wu2018}. 


Silicon-based anodes for Li-ion batteries represent a technologically relevant, stringent test case for XPS quantitative analysis. 
In fact, Si offers an exceptionally high theoretical capacity (4200 mAh g$^{-1}$), yet its practical use is limited by extreme volume changes and complex phase transformations during cycling, which generate large mechanical stresses and drive particle fracture, pulverization and loss of electrical contact\cite{wu2012,taiwo2017,muller2018,muller2020}.
These effects originate from intrinsically nonuniform (de)lithiation within individual particles\cite{domi2019,domi2020}, where Li concentration gradients induce sharp phase boundaries and anisotropic stresses, especially near interfaces between crystalline Li$_{3.75}$Si (c-Li$_{3.75}$Si) and amorphous Li$_x$Si phases\cite{obrovac2004,rohrer2013,gao2017}. Capturing these processes often demands \textit{operando} techniques with both surface sensitivity and chemical-state resolution.

Complementary \textit{in situ} X-ray diffraction (XRD)\cite{hatchard2004,li2007,misra2012}, transmission electron microscopy (TEM)\cite{wang2012,Liu2012,gu2013}, and nuclear magnetic resonance (NMR)\cite{key2009,ogata2014,kitada2019} have provided key insights into the structural evolution of Si anodes.
XPS adds an element- and chemical-state–resolved specificity, making it well suited to track near-surface lithiation-induced chemical changes. However, its quantitative interpretation has been hindered by the lack of a connection between XPS spectral features and Li concentration in disordered, nonstoichiometric phases that dominate Li–Si (de)lithiation. In this context, recent \textit{operando} XPS measurements on a-Si thin films have established systematic correlations between the Si $2p$ binding energy (BE) and Li content during electrochemical cycling, and identified pronounced spectral shifts as fingerprints of the the formation of c-Li$_{3.75}$Si upon deep lithiation and its amorphization during delithiation\cite{endo2022}. Turning these empirical correlations into quantitative, transferable stoichiometry–spectrum relationships nonetheless remains challenging.

Ab initio approaches based on density functional theory (DFT) can provide core-level BEs and XPS spectra with high accuracy, and are routinely used to support identification of experimental spectral features\cite{Smith2016,trinh2018}. However, these simulations would become prohibitively expensive for the structurally complex or disordered materials typical of battery applications \cite{D1TA01218E,Takatani_1991,Saito_2016,WANG2023122581,TATSUMISAGO2012342}, considering both the large supercells required to model the systems and the statistics required to sample the broad distribution of local environments, each of which needs to be treated separately (i.e., one calculation for each non-equivalent atomic environment). 
This type of limitations has motivated the impressive rise of machine-learning (ML) strategies, which can approximate DFT-level predictions at far lower cost~\cite{Schmidt2019,doi:10.1126/sciadv.1701816,deringer2021}. 
These methods can be used both to generate realistic atomistic ensembles through ML interatomic potentials (MLIP) \cite{MACEpaper,deringer2021,WANG2018178,zeng2023deepmdkit,Bartok2010,Behler2007,Smith2017,schutt2022schnetpack,Rupp2012,Butler2018,fourGenBeheler,Unke2019,Batzner_NatCommun_2022_v13_p2453,Fan2022,deepkit3}, which support molecular dynamics (MD) simulations of large systems over extended timescales \cite{LinfengWater2021,ImbalzanoGaAs2021,Natasha2021, Gigli2024,Staacke2022,Deringer2017,Bartok2018,maresca2018screw,Deringer2020,DeringerCaroCsanyi,Rosenbrock2021,Malosso2022,Tisi2021,Tisi2024,GigliLPS}, and to predict target properties directly from local structural descriptors \cite{deepwannier,APTpaper,penfold2024,aarva2019,aarva2021}.

Here, we combine an uncertainty-aware ML surrogate model for predicting Li $1s$ and Si $2p$ core-level BEs with large-scale MD simulations for appropriately sampling Li$_x$Si structures across a wide range of compositions. The surrogate model is trained on DFT reference calculations and provides calibrated uncertainty estimates, enabling robust predictions for both crystalline and disordered systems. To generate physically relevant configurations, we employ MD and grand-canonical Monte Carlo (GCMC) simulations driven by a neuro-evolution potential (NEP)\cite{Fan2022}. Using the resulting ensembles, we construct stoichiometry maps that directly relate XPS peak positions to Li content and structural evolution during delithiation. The resulting trends are consistent with available \textit{operando} and \textit{ex situ} experiments and reveal clear spectroscopic signatures of the crystalline-to-amorphous transition in Li-Si, offering a quantitative interpretation of XPS data from Si-based electrodes under realistic operating conditions.

\section{Methods}\label{sec:Methods}

\subsection{Core-level binding energies from DFT}

The core-level binding energies (BE) were computed using 
a pseudopotential-based total-energy difference 
method~\cite{Cavigliasso1999}, using Quantum ESPRESSO~\cite{Giannozzi2009,Giannozzi2017,Giannozzi2020} as DFT engine.
Within the excited core-hole (XCH)~\cite{prendergast2006} approximation, the BE of a core state $n$ for a target atom $a$ is computed as the difference between the total energy of the excited state $E^*$, with an electron removed from the core and placed in valence, and the ground state total energy $E^0$:
\begin{equation}\label{BE}
\text{BE}_{a,n} = E^{*}(\eta_{c,n}-1,\eta_v+1)-E^{0}(\eta_{c,n},\eta_v)+
\Delta E_\text{corr} ,
\end{equation}
where $\eta_{c,n}$ and $\eta_v$ are the populations of the core state $n$ and valence, respectively. The correction term $\Delta E_\text{corr}$ contains the relaxation effects of core electrons in the presence of a core-hole~\cite{mizoguchi2009} for a given exchange-correlation functional~\cite{walter2016} as embedded in the core-hole pseudopotential, making the computed BE comparable to experiments. For core states with non-zero orbital angular momentum ($l\neq0$), the spin-orbit coupling (SOC) contribution was incorporated through the following empirical correction

\begin{equation}\label{BE_SOC}
\text{BE}_{a,n,l,j} = \text{BE}_{a,n}+ (-1)^{j-l+\frac{3}{2}}\frac{\left|\Delta^{l}_\text{SOC}\right|}{2l+1}\left[j(j+1)-l(l+1)-\frac{3}{4}\right],
\end{equation}
where $j$ is one of the two angular momentum values for a given $l$, and $\left|\Delta^{l}_\text{SOC}\right|$ is the absolute experimental energy split between them.

The final XPS spectrum for a species $\mathcal{A}$ and a core-level state $n$ was computed from the corresponding atomic contributions $\text{BE}_{a,n}$ as:
%
\begin{equation}\label{XPS}
S_{\mathcal{A},n}(E,\sigma,\Gamma) = \sum_{a\in\mathcal{A}} V(E-\text{BE}_{a,n},\sigma,\Gamma),
\end{equation}
where $a$ runs over all atoms of species $\mathcal{A}$, and $V(.)$ is the empirical line
broadening function centered at BE$_{a,n}$ and represented by the convolution of the Gaussian and Lorentzian distributions, known as the Voigt profile\cite{voigt1912}. The Gaussian broadening $\sigma$ and the Lorentzian broadening $\Gamma$ account for the experimental resolution and the lifetime of the core-hole, respectively.

DFT calculations were performed using ultrasoft pseudopotentials with the PBE exchange correlation functional. Core-hole pseudopotentials were generated to describe excited atoms for the simulation of Li $1s$ and Si $2p$ BEs, whereas pseudopotentials from the SSSP library v1.1~\cite{prandini2018}  were used to describe ground state atoms. The energy cut-offs for the kinetic energy of the wave functions and the charge density were set to 100 Ry and 800 Ry, respectively. A supercell approach was employed to minimize interactions between excited atoms and their periodic images, maintaining a minimum distance of 8 {\AA} between replicas. The Brillouin zone corresponding to the supercell was sampled using a Monkhorst-Pack $k$-point mesh with a grid spacing of 0.25~\AA{}$^{-1}$.

\subsection{Machine learning model and uncertainty estimation}
The BE of a target atom was predicted using kernel ridge regression (KRR)\cite{murphy2012}, taking its local atomic environment as input, as follows:
\begin{equation}\label{RBF_kernel}
\hat{y}_{*}=\sum_{i=1}^N \alpha_i k(\mathbf{X_*},\mathbf{X}_i)=\sum_{i=1}^N \alpha_i\exp{-\gamma||\mathbf{X_*}-\mathbf{X}_i||^2},
\end{equation}
where $\hat{y}_{*}$ is the BE prediction, $\mathbf{X_*}$ and $\mathbf{X}_i$ are equivariant feature vectors, representing the local atomic environment of the target atom and the training samples, respectively. $k(\mathbf{\cdot},\mathbf{\cdot})$ denotes the kernel function, which computes the distance between feature vectors and, in this work, corresponds to an isotropic Gaussian kernel with a length scale parameter $\ell=1/\sqrt{2\gamma}$. The coefficients $\alpha_i$ are the dual variables computed as:
\begin{equation}\label{alphas}
\pmb{\alpha}=\left(\mathbf{K}+\lambda\mathbf{I}\right)^{-1}\mathbf{y},
\end{equation}
where $\mathbf{K}=k(\mathbf{X}_i,\mathbf{X}_j)$ is the kernel matrix, $\lambda>0$ is the regularization strength, penalizing large values of the regression weights, and $\mathbf{y}$ is the vector of the training target BEs.

In order to endow the ML predictions with an associated uncertainty, we have constructed a committee of $M$ equivalent models $\hat{y}_*^{(m)}$, by subsampling the training set \cite{musil2019fast}. Within this framework, it is straightforward to propagate the uncertainty to quantities derived from the original predictions \cite{grasselli2025}. The ensemble average for a quantity with functional dependence on the predictions, $\hat z_*=f(\hat{y}_*)$, is defined as 

\begin{equation}\label{ensemble_avg}
\hat{\overline{z}}_*=\frac{1}{M}\sum_{m=1}^M \hat{z}_*^{(m)},
\end{equation}
 with the following sample standard deviation 

\begin{equation}\label{sd}
s_z(\mathbf{X_*})=\left(\frac{1}{M-1}\sum_{m=1}^{M}{\left|\hat z_*^{(m)}-\hat{\overline{z}}_*\right|^2}\right)^{1/2},
\end{equation}
where $\hat z_*^{(m)}=f(\hat{y}_*^{(m)})$ were evaluated across the ensemble. To account for finite-ensemble effects, the standard deviation was calibrated as $s_z^{(\text{calib.})}(\mathbf{X_*})=\kappa_z s_z(\mathbf{X_*})$, using a scaling factor $\kappa_z$,  defined for small $M$ as\cite{imbalzano2021uncertainty,grasselli2025}

\begin{equation}\label{sf}
\kappa_z^2=-\frac{1}{M}+\frac{M-3}{M-1}\frac{1}{N_{\text{val}}}\sum_{i=1}^{N_{\text{val}}}{\frac{\left|{z}_i-{\hat{\overline{z}}}_i\right|^2}{s_z^2(\mathbf{X}_i)}},
\end{equation}
where ${{z}}_i$ denotes the reference value (ground-truth) evaluated on the validation dataset $\left\{\mathbf{X}_i\right\}_{i=1}^{N_{\text{val}}}$.

\subsection{Local Atomic Environment Descriptor}

Local atomic environments were described using the smooth overlap of atomic positions (SOAP) power spectrum\cite{bartok2013}, as implemented in librascal\cite{musil2021}. The atom-centered density was expanded using the Gaussian-type orbital (GTO) basis with $n_{max}=8$ and spherical harmonics with $l_{max}=6$ for the radial and angular parts, respectively. A maximum cut-off radius of 8 \AA~was applied, such that only atoms within this distance were included in the expansion. Their contributions were smoothly weighted using radial scaling, defined by the function $f_{cut}(r) ={[1+(r/r_{0})^4]}^{-1}$, with characteristic distance $r_0$. The SOAP similarity between two atomic environments $\mathbf{X}_i$ and $\mathbf{X}_i$ was measured using the SOAP kernel $\left(\mathbf{X}_i\cdot\mathbf{X}_j\right)^\zeta$, with $\zeta=4$.


\subsection{Dataset and Farthest Point Sampling}

We took advantage of an available dataset containing $\sim$62000 structures of Li$_x$Si\cite{Fu2023}, which was originally developed to train a deep potential (DP)\cite{WANG2018178,zhang2018} model for Li–Si battery anodes. The dataset includes ground-state crystalline Li, Si, and Li$_x$Si phases ($0 < x < 4.5$) from the Materials Project\cite{jain2020}, expanded through DFT optimization at the PBE level and \textit{ab initio} MD sampling, as well as amorphous a-Li$_x$Si configurations generated via iterative DeePMD and grand canonical Monte Carlo approaches, as detailed in Ref.~\citenum{Fu2023}. 

From the original dataset, two disjoint subsets were constructed for training and validation. For the training set, data sub-selection was performed to ensure uniform sampling across the structures while reducing the computational effort related to DFT-based BE calculations. This was done by sorting the dataset in descending order of structural diversity using Farthest Point Sampling \cite{Eldar1997} (FPS). The algorithm was initialized with the first structure in the dataset, and in subsequent iterations, it selected the next sample based on the maximum distance criterion, as

\begin{equation}\label{FPS}
\langle\mathbf{X}\rangle_k=\underset{\langle\mathbf{X}\rangle_k\notin\mathcal{S}}{\operatorname{\text{argmax }}}\left(\underset{\langle\mathbf{X}\rangle_j\in\mathcal{S}}{\operatorname{\text{min }}}||\langle\mathbf{X}\rangle_k-\langle\mathbf{X}\rangle_j||^2\right),
\end{equation}
where $\mathcal{S}$ denotes the set of structures selected in the previous steps, and $\langle\mathbf{X}\rangle_k$ is the $k$-th structure-averaged SOAP vector, constructed using $r_{0}=6$ \AA\space and an atomic density smoothing $\sigma_{a}=0.3$ \AA~ for the GTO radial basis. The size of the training set was chosen to ensure that the mean absolute error (MAE) of the model remained below the typical experimental uncertainty for the BE resolution in XPS, i.e., 0.1 eV\cite{Seah2022}. 

\subsection{MD and GCMC Computational Details}

The melt-quench-anneal (MQA) protocol employed to generate near-equilibrium amorphous Li$_x$Si structures involves an initial melting phase from 2500 K and to 1800 K over 200 ps, followed by three quenching stages: (i) from 1500 K to 1250 K at a rate of $10^{13}$ Ks$^{-1}$,  (ii) from 1250 K to 1050 K at a rate of 10$^{11}$ Ks$^{-1}$, and (iii) from 1050 K to 500 K at a rate of $10^{13}$ Ks$^{-1}$. The melting and quenching phases were simulated using the isobaric-isothermal ($NPT$) ensemble, at $P=0$ bar, by coupling the system (assumed here isotropic) with the stochastic cell rescaling (SCR) barostat\cite{bernetti2020} and the stochastic velocity rescaling (SVR) thermostat\cite{bussi2007}. After quenching, the atomic positions were further relaxed using the steepest descent method to achieve mechanical equilibrium with a force tolerance of $10^{-6}$ eV\AA$^{-1}$. The corresponding simulations were performed with \textsc{GPUMD}\cite{Fan2022} using a Li-Si neuroevolution potential\cite{Fan2022} (NEP) from Ref.~\citenum{Pegolo2024}.

To investigate the early stages of the delithiation process, grand canonical Monte Carlo (GCMC) simulations\cite{Frenkel2023} were carried out starting from different crystalline phases of Li silicides having different initial Li concentration. For each structure, two particle exchanges and one displacement attempt were performed every 50 steps of molecular dynamics (MD) in the $NPT$ ensemble, with $P=0$ bar and $T=300$ K. The chemical potential ($\mu$) was dynamically changed to favor the delithiation process with an acceptance ratio of 44$\%$. The simulations were conducted using \textsc{LAMMPS}\cite{Thompson2022} compiled with the PWMLFF package for the NEP-LAMMPS interface. MD simulations were performed using a hybrid / overlay setup that combines NEP and Ziegler-Biersack-Littmark\cite{Ziegler1985} (ZBL) potentials to avoid nonphysical bond lengths during volume contraction in the delithiation process. 

\section{Results and discussion}\label{sec:Results}

\subsection{Model Training and Validation}


\begin{figure}
\centering
    \includegraphics[width=0.8\textwidth]{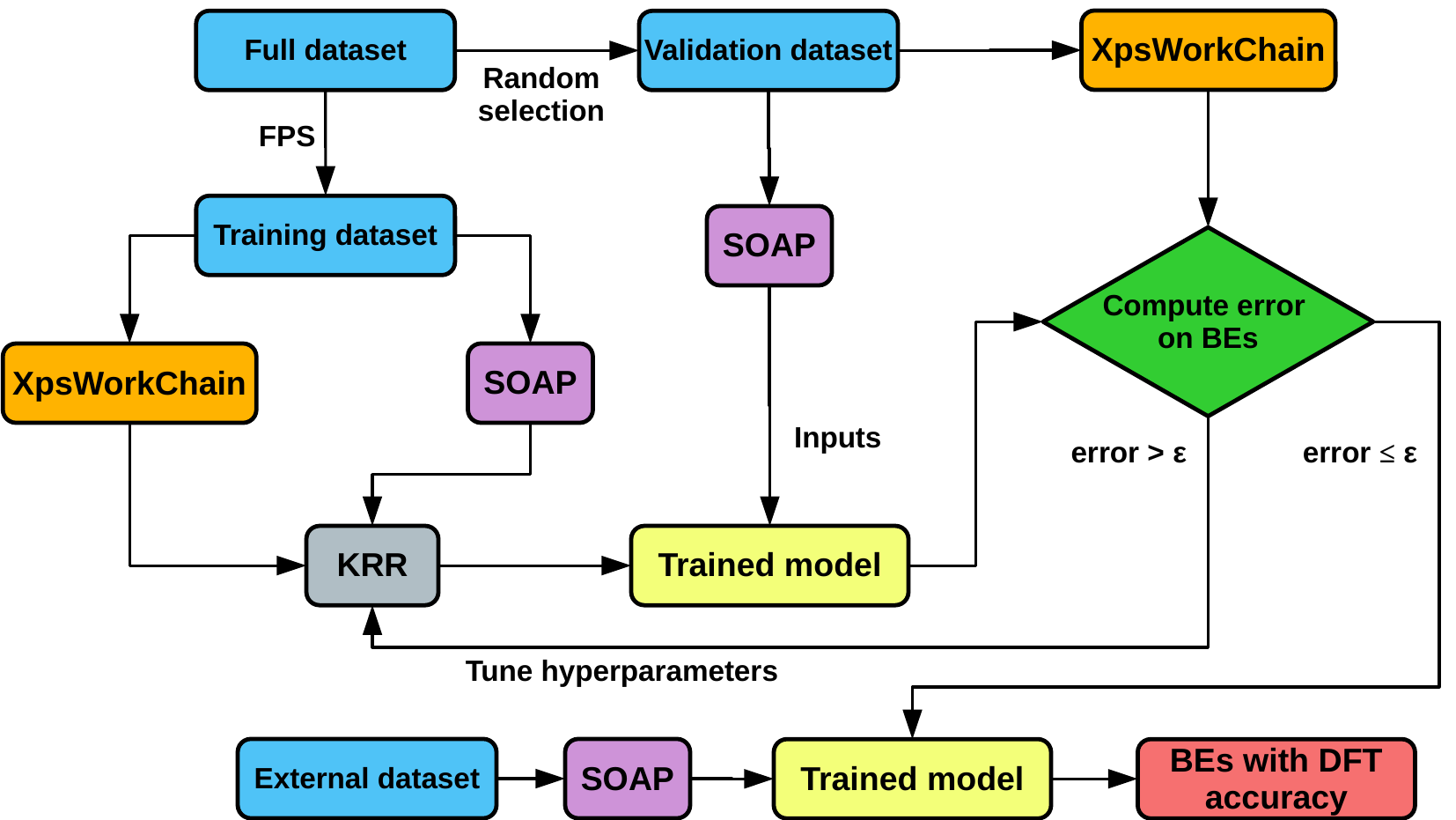}
\caption{Schematic of the machine-learning workflow for BE prediction, including data generation, training, and validation steps.}
\label{fig:training_diagram}
\end{figure}

The general scheme of the training and validation process is depicted in \autoref{fig:training_diagram}, where the individual steps are described in the Methods Section. 
Starting from the full dataset of Li$_x$Si structures, the final training dataset obtained by FPS includes 245 structures, comprising $\sim$7800 Li and $\sim$7800 Si atomic environments. The validation set -- randomly selected from the remaining structures -- contains instead 55 structures, with $\sim$1900 Li and $\sim$1900 Si atomic environments.

For each selected structure in both the training and validation sets, the DFT BEs for Li $1s$ and Si $2p$ were computed by exploiting  the \texttt{XpsWorkChain} of the \texttt{aiida-qe-xspec} plugin~\cite{aiida_qe_xspec}, an automated Python workflow based on AiiDA \textsc{Quantum ESPRESSO}~\cite{huber2020,uhrin2021} that implements the calculation XPS BEs as described in the Methods Section. The \texttt{XpsWorkChain} has been interfaced seamlessly with our Python workflow for the training and validation of the KRR models, enabling consistency and high-throughput generation of the training data as well as full data provenance as provided by AiiDA. The training of the KRR models was performed with the scikit-learn Python library \cite{pedregosa2011}, using a grid search over the hyperparameters space $(\lambda,\gamma)$ to find the optimal values that minimize the MAE in a five-fold cross-validation configuration. A total of 12 models were trained for each of the two core-levels, covering all combinations of characteristic distances ${r_{0}}=2.5 , 3.5, 4.5, 5.5$ \AA~and atomic density smoothing values ${\sigma_a}=0.3, 0.4, 0.5$ \AA~used for the SOAP descriptors (see Figure S1, Supporting Information). The models with the best bias-variance trade-off, defined here as those that maximize the coefficient of determination $R^2$ on the validation set while minimizing the difference between the training and validation $R^2$, were selected as the optimal models. 


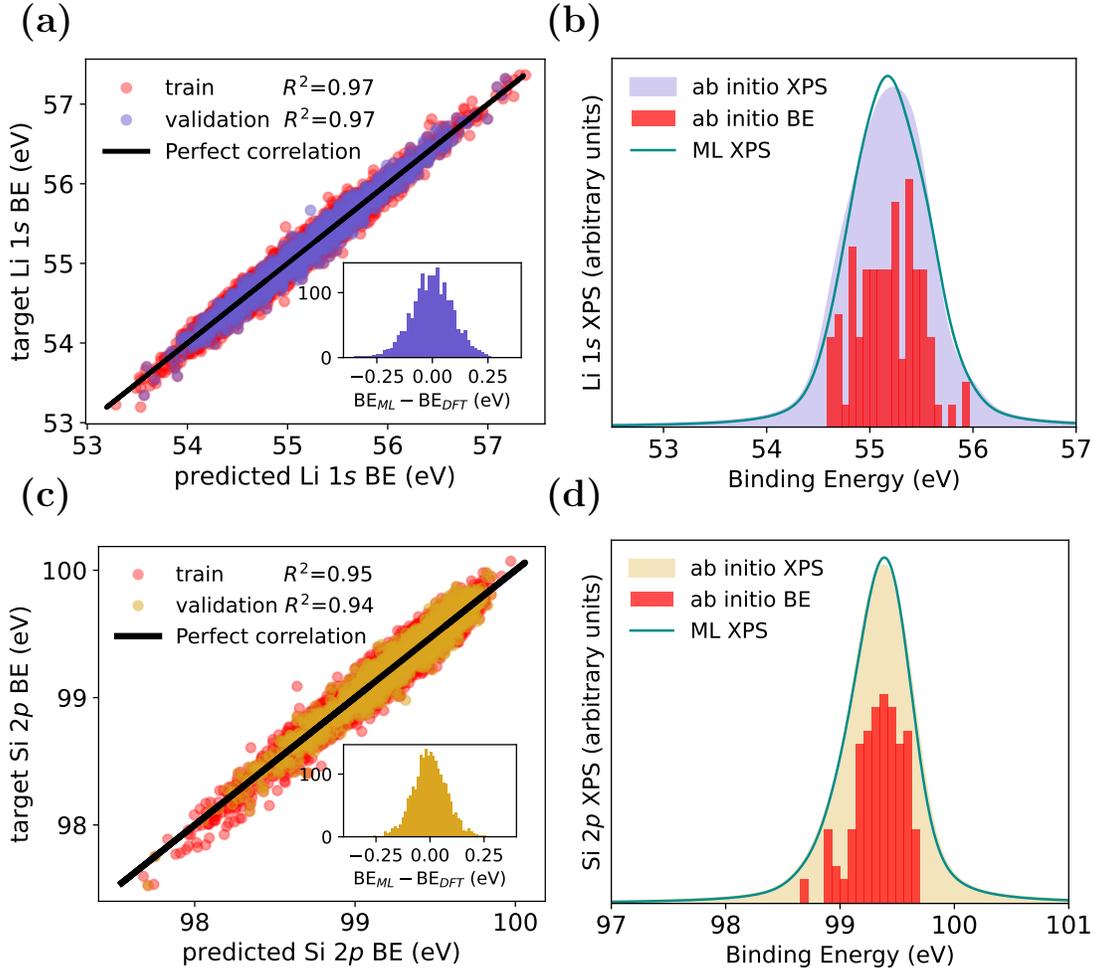
\begin{figure}[htbp]
    \centering

    \begin{subfigure}[b]{0.45\textwidth}
        \includegraphics[width=\textwidth]{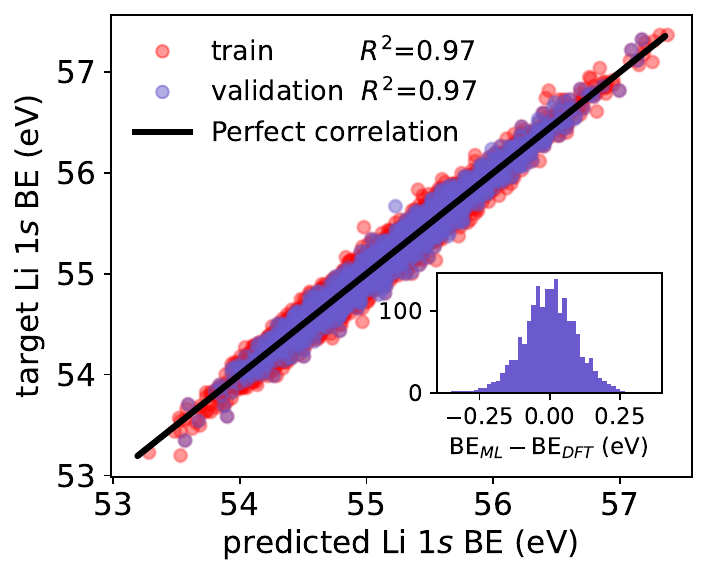}
        \caption*{}
        \label{fig:1}
    \end{subfigure}
    \hspace{-0.1cm}
    \begin{subfigure}[b]{0.43\textwidth}
        \includegraphics[width=\textwidth]{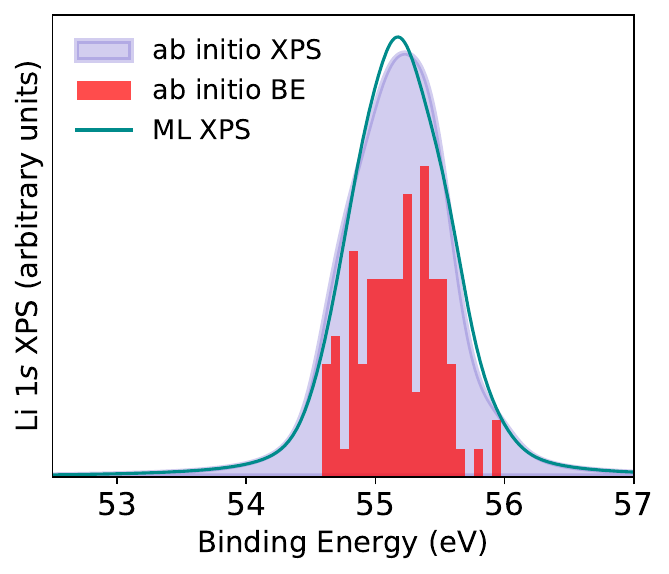}
        \caption*{}
        \label{fig:2}
    \end{subfigure}

    \vspace{-0.4cm}

    \begin{subfigure}[b]{0.45\textwidth}
        \includegraphics[width=\textwidth]{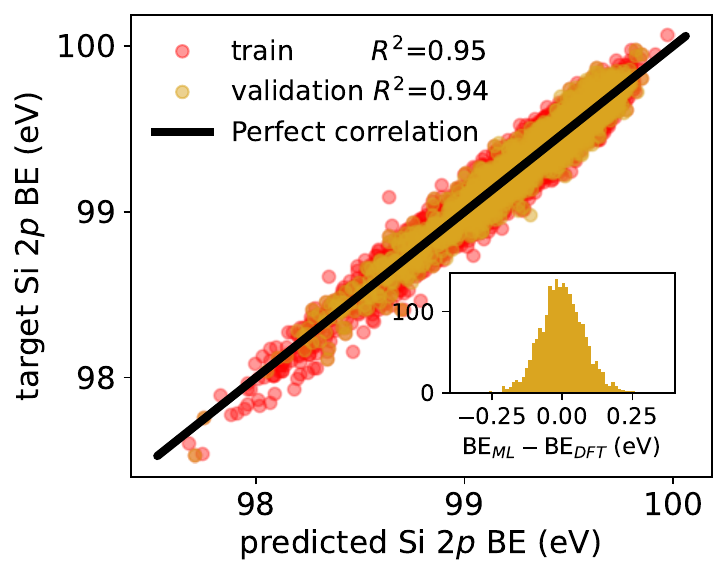}
        \caption*{}
        \label{fig:3}
    \end{subfigure}
    \hspace{-0.1cm}
    \begin{subfigure}[b]{0.43\textwidth}
        \includegraphics[width=\textwidth]{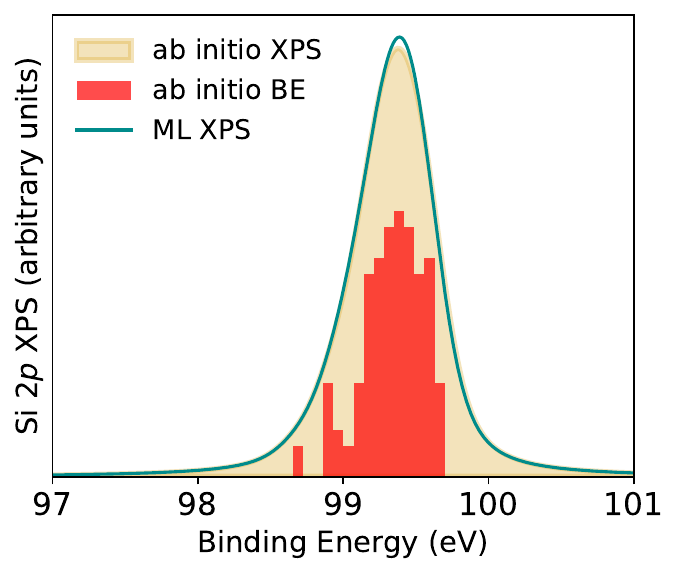}
        \caption*{}
        \label{fig:4}
    \end{subfigure}
    
 \setlength{\unitlength}{1cm}
\begin{picture}(0,0)
    \put(-7.0,13.8){\large\textbf{(a)}}   
    \put(0.0,13.8){\large\textbf{(b)}}
    \put(-7.0,7.5){\large\textbf{(c)}}   
    \put(0.0,7.5){\large\textbf{(d)}}
\end{picture}
    \vspace{-0.7cm}
    \caption{(Left) Performance of the Gaussian KRR models corresponding to the Li $1s$ (a) and Si $2p$ (c) for the Li$_x$Si train and validation sets. The determination coefficient $R^2$ for the relation between ML predictions and \textit{ab initio} DFT (target) is reported in the legend. The histogram for the difference between BEs obtained
with ML and \textit{ab initio} DFT is shown as an inset. (Right) Model benchmarking against DFT XPS obtained for an out-of-sample \ce{Li$_x$Si} structure from Ref.\citenum{chen2021}, the red histograms account for the DFT BEs of Li $1s$ (b) and Si $2p$ (d). The colored areas (curves in darkcyan) represent the XPS spectra obtained by adding Voigt profiles centered on the DFT (ML predicted) BEs.}
    \label{fig:panel_train_validation}
\end{figure}

\autoref{fig:panel_train_validation} presents the training results for the optimal models predicting the Li $1s$ and Si $2p$ BEs in the Li$_x$Si dataset. In the left panels, we present the correlation plots for the Li $1s$ (a) and Si $2p$ (c) BEs between the best model predictions and the \textit{ab initio} DFT reference for both the training and the validation set. These models achieved $R^2$ values of 0.97 and 0.94 for Li $1s$ and Si $2p$, respectively. The corresponding hyperparameters, along with the MAE and root mean square error (RMSE) of the best models, are reported in Supporting Information (see Table S1). In order to test their generalization capability, the models were benchmarked against DFT XPS data for an out-of-sample \ce{Li$_x$Si} structure from Ref.~\citenum{chen2021}. The chosen structure contains 224 atoms, i.e., well beyond the average and maximum size of the training set (64 and 152 atoms, respectively). The results of the benchmark are reported on the right panels of \autoref{fig:panel_train_validation}, where the red histograms represent the DFT BEs for the Li $1s$ (b) and Si $2p$ (d) core-levels, while the shaded areas and the dark cyan curves correspond to the XPS spectra reconstructed as the sum of Voigt profiles centered on the BEs obtained by DFT and the ML models, respectively. The agreement between the predicted spectra and the reference DFT data demonstrates the transferability of the models beyond the training set. 

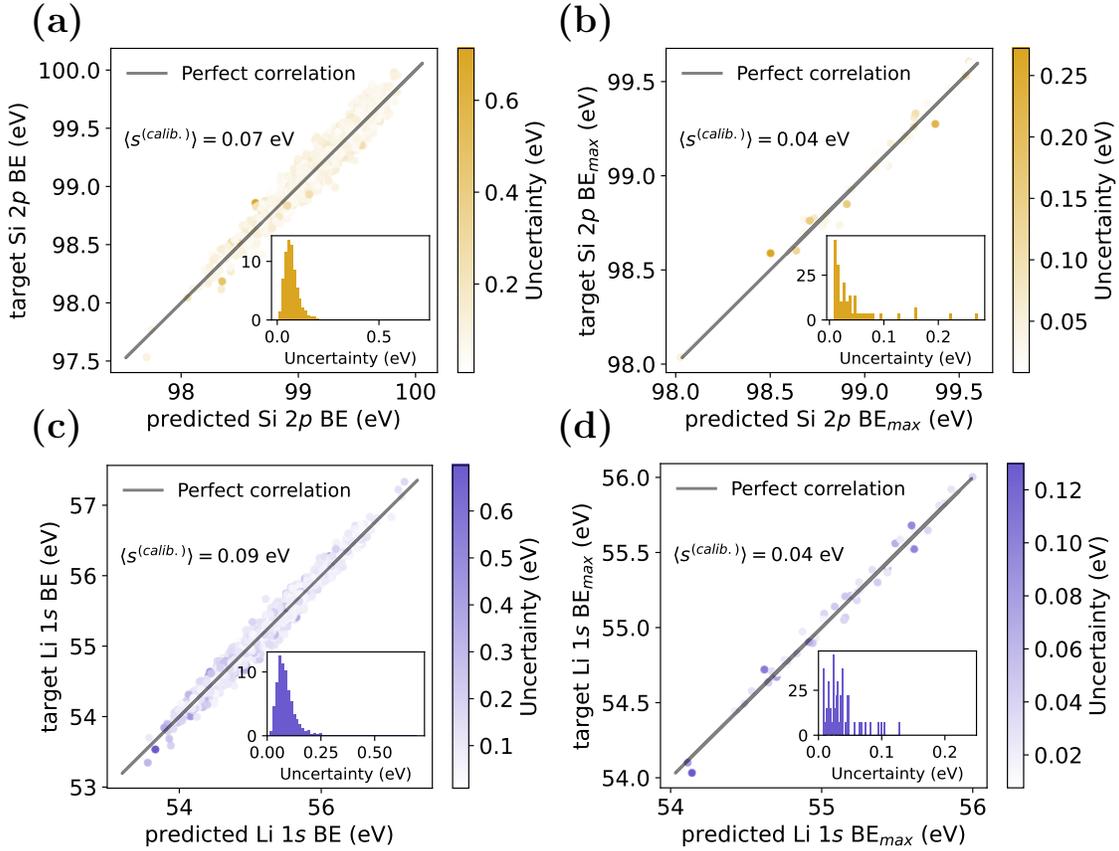
\begin{figure}[htbp]
    \centering

    \begin{subfigure}[b]{0.45\textwidth}
        \includegraphics[width=\textwidth]{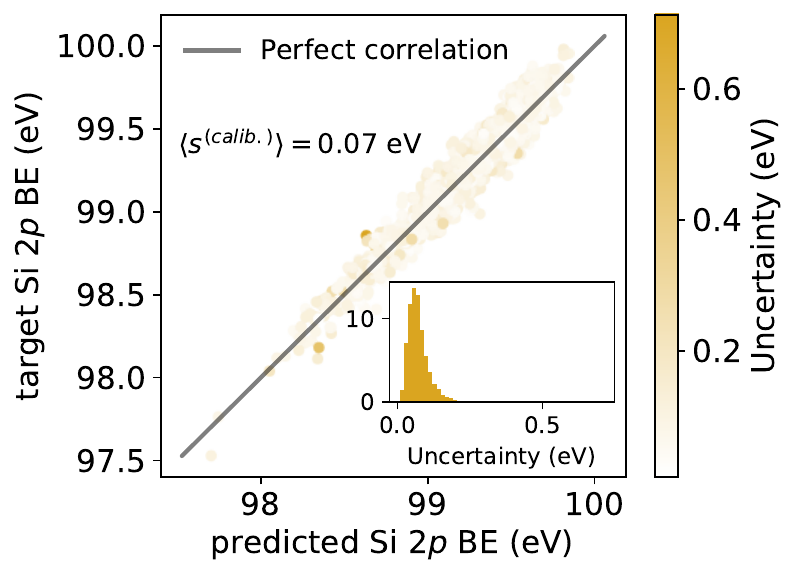}
        \caption*{}
        \label{fig:310}
    \end{subfigure}
    \begin{subfigure}[b]{0.45\textwidth}
        \includegraphics[width=\textwidth]{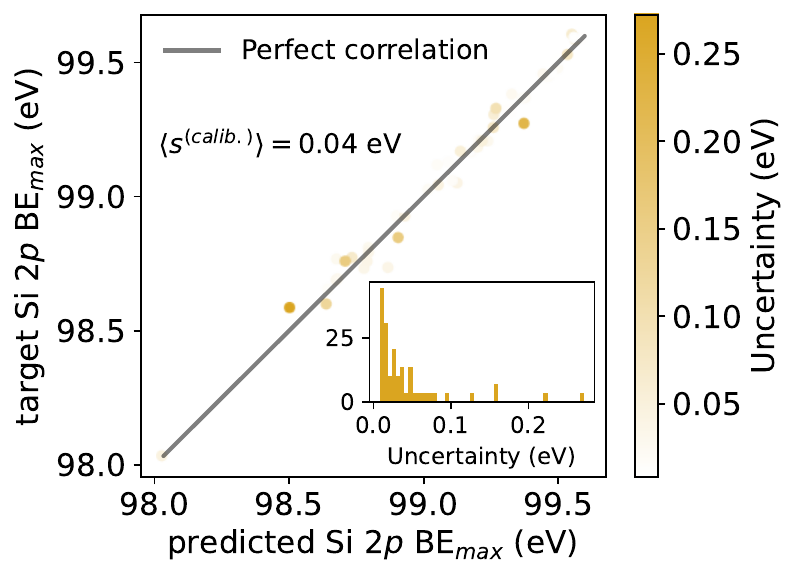}
        \caption*{}
        \label{fig:320}
    \end{subfigure}
        \begin{subfigure}[b]{0.42\textwidth}
        \vspace{-0.6cm}
        \hspace{0.15cm}
        \includegraphics[width=\textwidth]{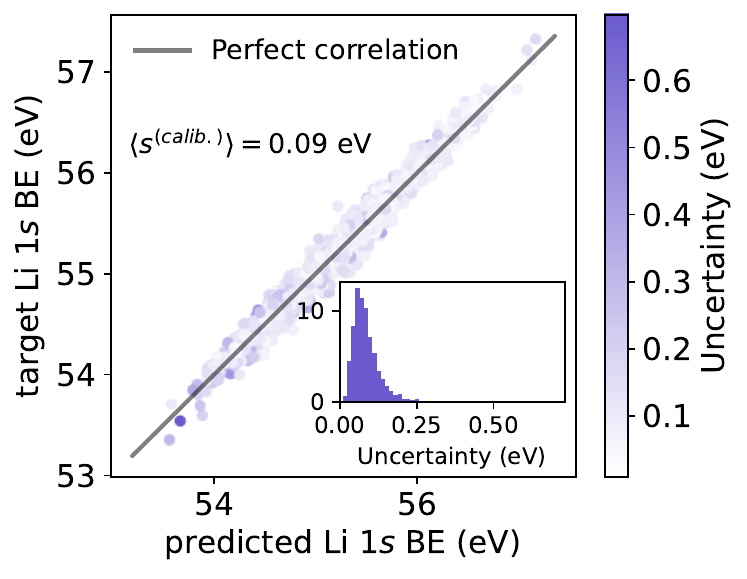}
        \caption*{}
        \label{fig:330}
    \end{subfigure}
    \begin{subfigure}[b]{0.45\textwidth}
    \vspace{-0.6cm}
    \hspace{0.15cm}
        \includegraphics[width=\textwidth]{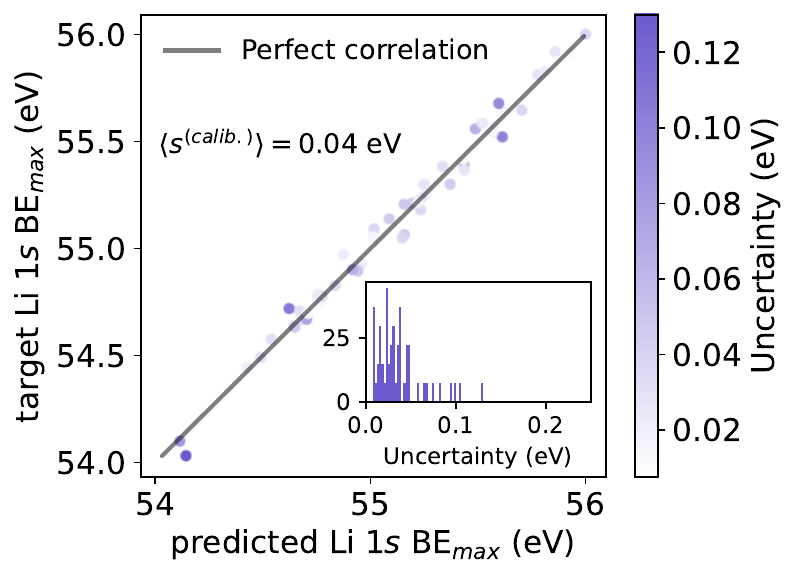}
        \caption*{}
        \label{fig:340}
    \end{subfigure}
    \setlength{\unitlength}{1cm}
    
\begin{picture}(0,0)
    \put(-7.0,12.2){\large\textbf{(a)}}   
    \put(0.0,12.2){\large\textbf{(b)}}
    \put(-7.0,6.8){\large\textbf{(c)}}   
    \put(0.0,6.8){\large\textbf{(d)}}
\end{picture}
    \vspace{-0.7cm}
    \caption{Uncertainty estimates for the BE (a,c) and BE$_\text{max}$ (b,d) predictions on the validation set. (a) and (b) panels correspond to the Si $2p$ level, while (c) and (d) show results for the Li $1s$ level. Each plot reports the average calibrated uncertainty $\langle s^{(calib.)}\rangle$, and the insets display histograms of the corresponding uncertainty distributions.}
    \label{fig:uncertainty}
\end{figure}

Uncertainty estimates for the BEs and for $\text{BE}_\text{max}=\underset{E}{\operatorname{argmax}}\left[S_{\mathcal{A},n}(E,\sigma,\Gamma)\right]$, derived from the model predictions, are shown in \autoref{fig:uncertainty}. Overall, the calibrated uncertainties predicted for the BEs (a,c) are consistent with typical XPS experimental resolutions, on average falling below 0.1 eV. For BE$_\text{max}$ (b,d), the quantity typically monitored in operando experiments, the propagated uncertainty is even smaller, further confirming the reliability of the model predictions. 

\subsection{Comparison with Experimental Data}

The experimental reference data in this work are drawn from two XPS studies on a-Si thin films as model electrodes for Li-ion batteries. In the first study, a-Si thin films deposited on Cu were cycled in a liquid-electrolyte cell, and \textit{ex situ} XPS coupled with scanning electron microscopy (SEM) was used to track the potential-dependent evolution of both the Si electrode and SEI\cite{ferraresi2016}. In the second study, a-Si films were sputter-deposited onto a solid electrolyte \ce{Li_{6.6}La3Zr_{1.6}Ta_{0.4}O12} and subjected to electrochemical lithiation and delithiation, which were monitored by \textit{operando} XPS. Upon deep lithiation and subsequent delithiation, an abrupt shift in the Si $2p$ BE was observed and attributed to the phase transition between c-\ce{Li_{3.75}Si} and a-Li$_x$Si, reflecting structural and electronic changes during cycling\cite{endo2022}. Collectively, these studies provide XPS benchmarks for the stoichiometric and structural evolution of Si-based electrodes upon cycling, against which ML-predicted BEs and spectral trends can be directly assessed.

Building on these experimental benchmarks, a first attempt to construct a stoichiometry map for the discharge of a Si electrode was carried out by applying the ML model to the complete Li$_x$Si dataset\cite{Fu2023} (see Figure S2, Supporting Information). The predicted BE$_\text{max}$ values show no clear dependence on Li concentration for the Li $1s$ core-level, whereas the Si $2p^{3/2}$ BE exhibits an upward trend with decreasing Li content, as expected upon delithiation. Moreover, the standard deviation of BE$_\text{max}$ in the dataset is relatively small at low Li concentrations, but increases substantially for $x > 1.5$. This is partly attributable to the diversity of local atomic environments in the dataset, beneficial for training a MLIP, but which includes motifs unlikely to occur under galvanostatic cycling conditions. In addition, amorphous structures at high Li concentrations are underrepresented, further contributing to the increased uncertainty. These observations highlight the need to generate additional, physically relevant structures, in particular a-Li$_x$Si configurations.

\subsubsection{Generation of amorphous Li$_x$Si structures}


To generate realistic Li$_x$Si amorphous structures, both near-equilibrium and compositionally driven configurations were prepared, following the protocols detailed in the Methods Section. The near-equilibrium amorphous structures were generated using the melt-quench-anneal (MQA) procedure described in Ref.~\citenum{Pegolo2024}, with stoichiometries ranging from $x$=0 to 4.4 in a 76000-atom supercell. 
The compositionally driven structures were instead generated using grand canonical Monte Carlo (GCMC) simulations\cite{Frenkel2023} coupled with MD to address the delithiation process. We considered as starting points different crystalline phases with variable initial compositions, namely c-\ce{Li_{3.75}Si}, (a-)c-\ce{Li_{4.4}Si}, c-\ce{Li_{3.25}Si} and c-\ce{Li_{1.7}Si}, using supercells containing 76000, 76000, 58752 and 27360 atoms, respectively. The metastable phase c-\ce{Li_{3.75}Si} is included as it is the key crystalline compound formed electrochemically when a-Li$_x$Si reaches a critical concentration\cite{hatchard2004,obrovac2004,li2007}, while the other crystalline phases, taken from the equilibrium Li-Si phase diagram\cite{van1985}, serve as structural references to characterize the evolution of Li$_x$Si. An amorphous configuration (a-\ce{Li_{4.4}Si}) was also studied to isolate the effects of structural order on the delithiation mechanism and the resulting XPS spectrum.

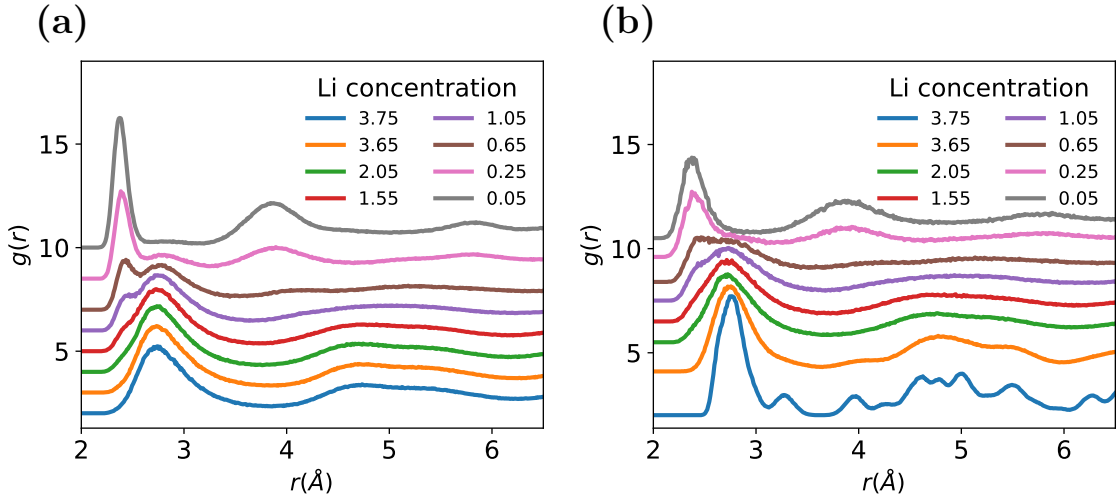
\begin{figure}[htbp]
    \centering
\vspace{0.5cm}
    \begin{subfigure}[b]{0.45\textwidth}
        \includegraphics[width=\textwidth]{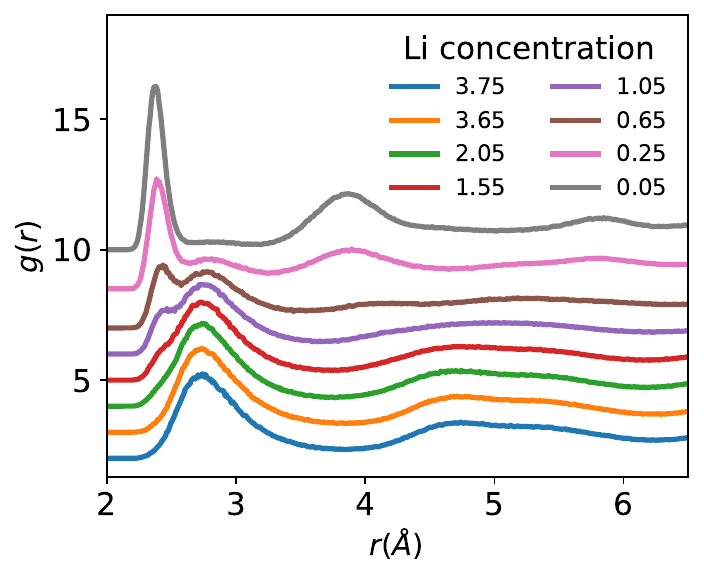}
        \caption*{}
        \label{fig:31}
    \end{subfigure}
    \hspace{-0.1cm}
    \begin{subfigure}[b]{0.45\textwidth}
        \includegraphics[width=\textwidth]{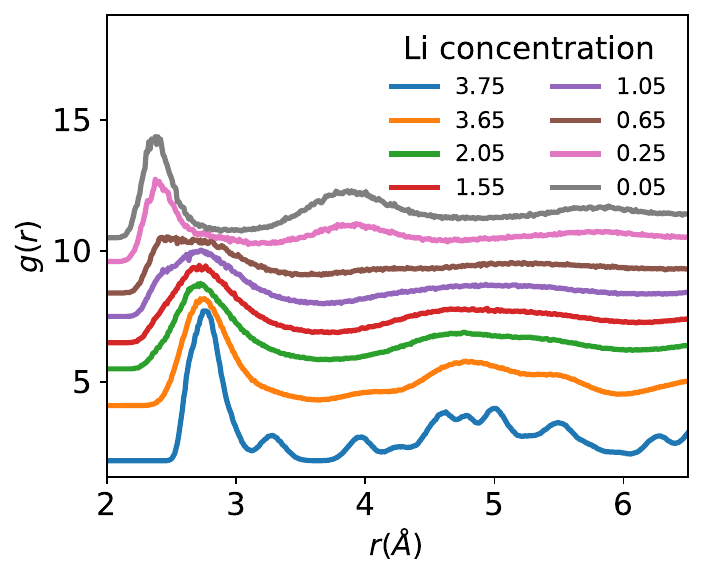}
        \caption*{}
        \label{fig:32}
    \end{subfigure}
    \setlength{\unitlength}{1cm}
    
\begin{picture}(0,0)
    \put(-7.0,7.5){\large\textbf{(a)}}   
    \put(0.5,7.5){\large\textbf{(b)}}
\end{picture}
    \vspace{-0.9cm}
    \caption{Radial distribution functions (RDFs) at 300 K of the structures generated using a melt-quench-anneal procedure (a) and a grand canonical Monte Carlo simulation starting from c-\ce{Li_{3.75}Si} (b).}
    \label{fig:RDF}
\end{figure}

In \autoref{fig:RDF}, the radial distribution functions (RDFs) of the generated structures are shown, comparing those obtained with the MQA procedure (a) and the GCMC-based delithiation protocol starting from c-\ce{Li_{3.75}Si} (b) at different Li concentrations. Both methods display broadened features and attenuation of long-range order, consistent with disordered atomic environments, except in the vicinity of $x=3.75$, where  the structures generated with GCMC show coexistence of crystalline and amorphous features. By combining MQA and GCMC, the dataset achieves broader coverage, with MQA providing amorphous environments across concentrations and GCMC capturing metastable or partially ordered motifs relevant to early delithiation.

Convergence tests were carried out on the BE values computed along the GCMC trajectories by considering the number of MD steps per MC move, the acceptance ratio and the system size (see Figure S3, Supporting Information). No significant changes were observed in the predicted BE values with respect to the first two parameters, while a dependence on the system size is observed, with convergence achieved only for systems containing at least 8000 Si atoms (within an initial \ce{Li_{3.75}Si} cell of 76000 atoms). These sizes clearly exceeds the practical limits of conventional DFT-based simulations, further supporting the suitability of the ML approach for efficiently predicting XPS-related properties in realistic Li–Si systems. 

\subsubsection{Li $1s$ spectrum and voltage profile during Li$_x$Si delithiation}

\begin{figure}[htbp]
    \centering
    \vspace{0.5cm}
\hspace{0.4cm}
    \begin{subfigure}[b]{0.45\textwidth}
        \includegraphics[width=\textwidth]{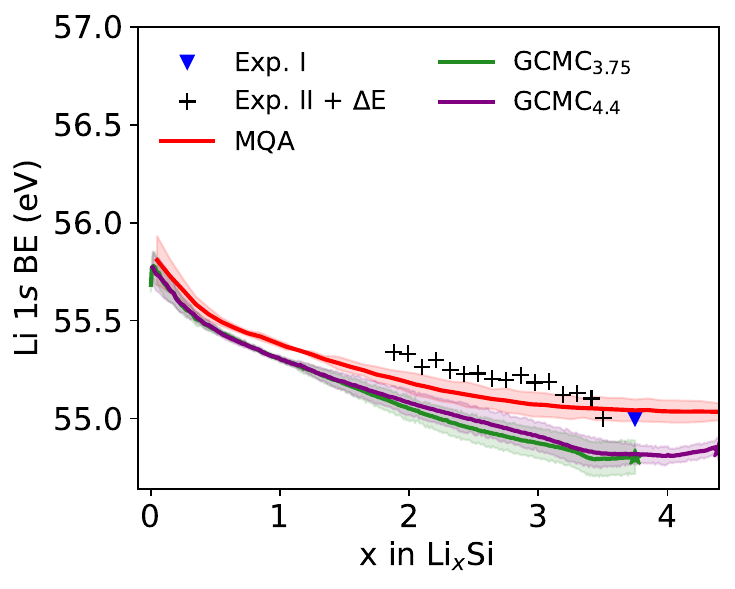}
        \caption*{}
        \label{fig:stoichiometry_map_Li_0}
    \end{subfigure}
    \hspace{-0.0cm}
    \begin{subfigure}[b]{0.45\textwidth}
        \includegraphics[width=\textwidth]{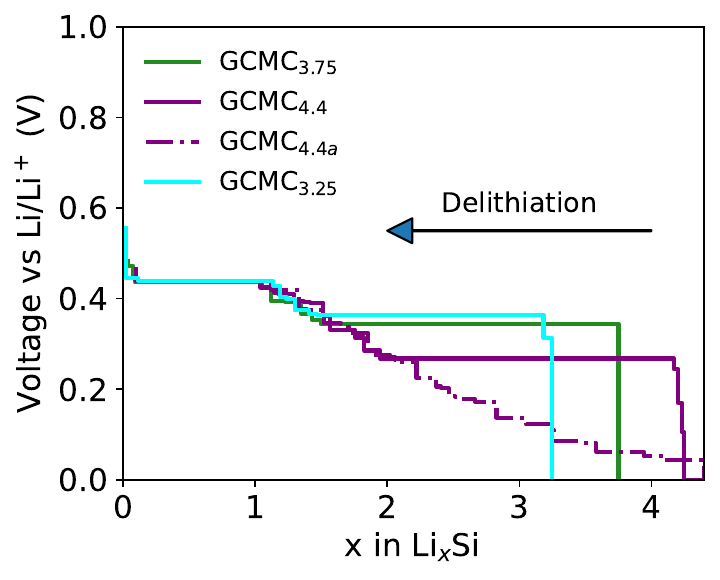}
        \caption*{}
        \label{fig:stoichiometry_map_V_x}
    \end{subfigure}

    \setlength{\unitlength}{1cm}
    \begin{picture}(0,0)
    \put(-6.5,7.5){\large\textbf{(a)}}   
    \put(1.0,7.5){\large\textbf{(b)}}
\end{picture}
    \vspace{-0.7cm}
    \caption{ (a) Stoichiometry map showing the peak position of Li $1s$  for Li$_x$Si structures generated using MQA (red) and GCMC simulations starting from c-\ce{Li_{3.75}Si} (green) and c-\ce{Li_{4.4}Si} (purple),  compared to experimental data corresponding to the delithiation of Si electrodes (black crosses and blue triangles). The stars represent the values for the initial structure in the GCMC simulations and the shaded regions correspond to the predicted uncertainties. (b) Delithiation voltage curves from GCMC simulations starting from c-\ce{Li_{3.75}Si} (green), c-\ce{Li_{4.4}Si} (purple), a-\ce{Li_{4.4}Si} (purple dashed) and c-\ce{Li_{3.25}Si} (cyan).}
    \label{fig:stoichiometry_map_Li}
\end{figure}
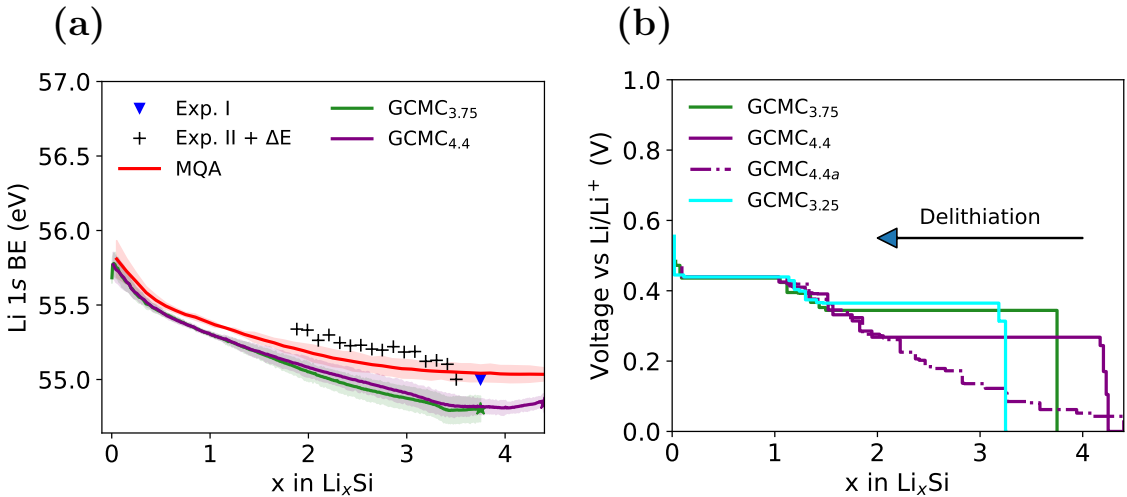

The ML model was applied to the Li$_x$Si structures generated using MQA and GCMC, and the resulting Li $1s$ BEs are compared in \autoref{fig:stoichiometry_map_Li}(a). The peak positions obtained from MQA structures (red solid line) and GCMC simulations starting from c-\ce{Li_{3.75}Si} (green solid line) and c-\ce{Li_{4.4}Si} (purple solid line) are consistent with experimental measurements of Si electrode delithiation, where the first dataset (black crosses) was extracted from Ref.~\citenum{endo2022} and the second dataset (blue triangles) from Ref.~\citenum{ferraresi2016}. To align the two experimental datasets, the energies of the first dataset (black crosses) were rigidly shifted by $\Delta E=2.86$ eV.  Details of the data extraction, spectral fitting, and alignment procedures are provided in the Supporting Information (Figure S4).
Both simulation protocols reproduce the experimentally observed downshift of the Li $1s$ binding energy with increasing Li concentration. However, the GCMC results are systematically shifted to slightly lower binding energies compared to MQA. The corresponding delithiation voltage curves, shown in \autoref{fig:stoichiometry_map_Li}(b), were obtained from the GCMC-generated structures, including those derived from c-\ce{Li_{3.75}Si} (green), c-\ce{Li_{4.4}Si} (purple), a-\ce{Li_{4.4}Si} (purple dashed), and c-\ce{Li_{3.25}Si} (cyan). These curves were constructed by neglecting the enthalpic and entropic contributions to the Gibbs free energy. Under this approximation, the voltage relative to Li/Li$^+$ for a transition between two phases of Li$_x$Si can be expressed as\cite{aydinol1997,chevrier2009,zhang2013}

\begin{equation}\label{voltage}
eV(x)\approx-\frac{E_f(x+\Delta x)-E_f(x)}{\Delta x},
\end{equation}
with

\begin{equation}\label{formation_E}
E_f(x)=E_{\text{Li}_{x}\text{Si}}-xE_\text{Li}-E_\text{Si},
\end{equation}
where $E_f$ denotes the formation energy per Si atom in eV, $E_{\text{Li}_{x}\text{Si}}$ is the total energy per Si atom of the Li$_x$Si phase, $E_\text{Li}$ the total energy per atom of BCC lithium, and $E_\text{Si}$ the total energy per atom of amorphous silicon. In \autoref{voltage}, the formation energy is evaluated at the vertices of the convex hull constructed from the $E_f(x)$ curve (see Figure S5, Supporting Information). The figure shows that delithiation of c-Li$_x$Si proceeds through a two-phase mechanism, manifested as a voltage plateau that connects the initial crystalline and final amorphous phases of Li$_x$Si. The estimated plateau voltages are 0.27 V for c-\ce{Li_{4.4}Si}, 0.34 V for c-\ce{Li_{3.75}Si}, and 0.36 V for c-\ce{Li_{3.25}Si}. For c-\ce{Li_{3.75}Si}, our computed plateau voltage falls within the lower end of experimentally reported values (0.38–0.45 V) \cite{ogata2014,Iaboni2016,gao2017,kitada2019,barman2020,woodard2021}, and is consistent with previous theoretical predictions of 0.27 V \cite{Fu2023}. These plateaus provide insight into the sequence in which the c-Li$_x$Si phases undergo delithiation. In contrast, the a-\ce{Li_{4.4}Si} begins delithiation at voltages substantially lower than those of the crystalline phases, indicating that lithium is preferentially extracted from the amorphous environments rather than from the crystalline ones.  

%

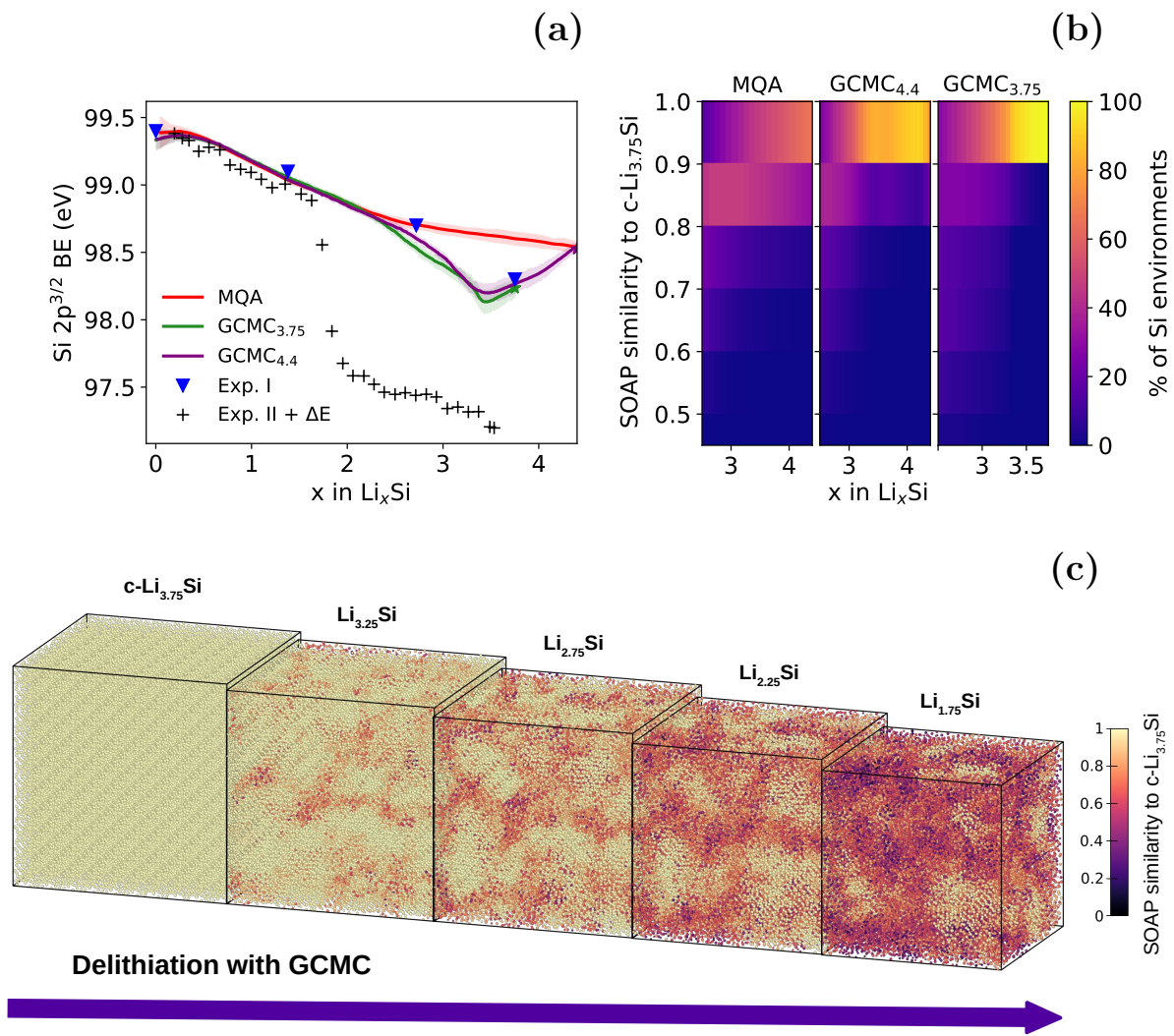
\begin{figure}[htbp]

    \begin{subfigure}[b]{0.45\textwidth}
        \includegraphics[width=\textwidth]{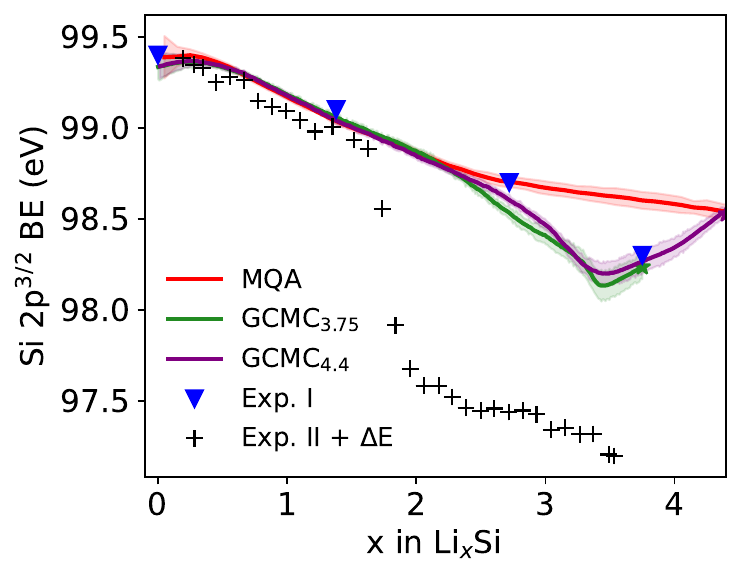}
        \caption*{}
        \label{fig:41}
    \end{subfigure}
    \hspace{-0.0cm}
    \begin{subfigure}[b]{0.47\textwidth}
        \includegraphics[width=\textwidth]{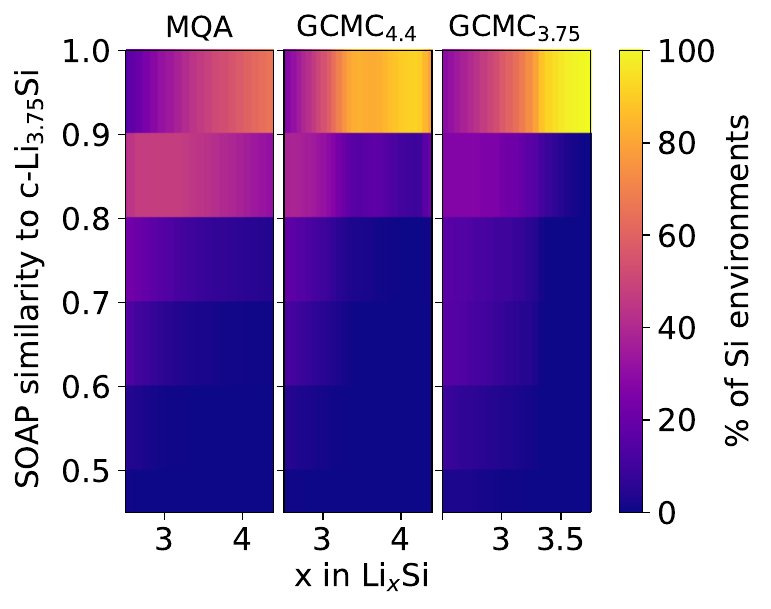}
        \caption*{}
        \label{fig:42}
    \end{subfigure}
    \centering
   \hspace{0.4cm}
        \begin{subfigure}[b]{1.0\textwidth}

        \includegraphics[width=\textwidth]{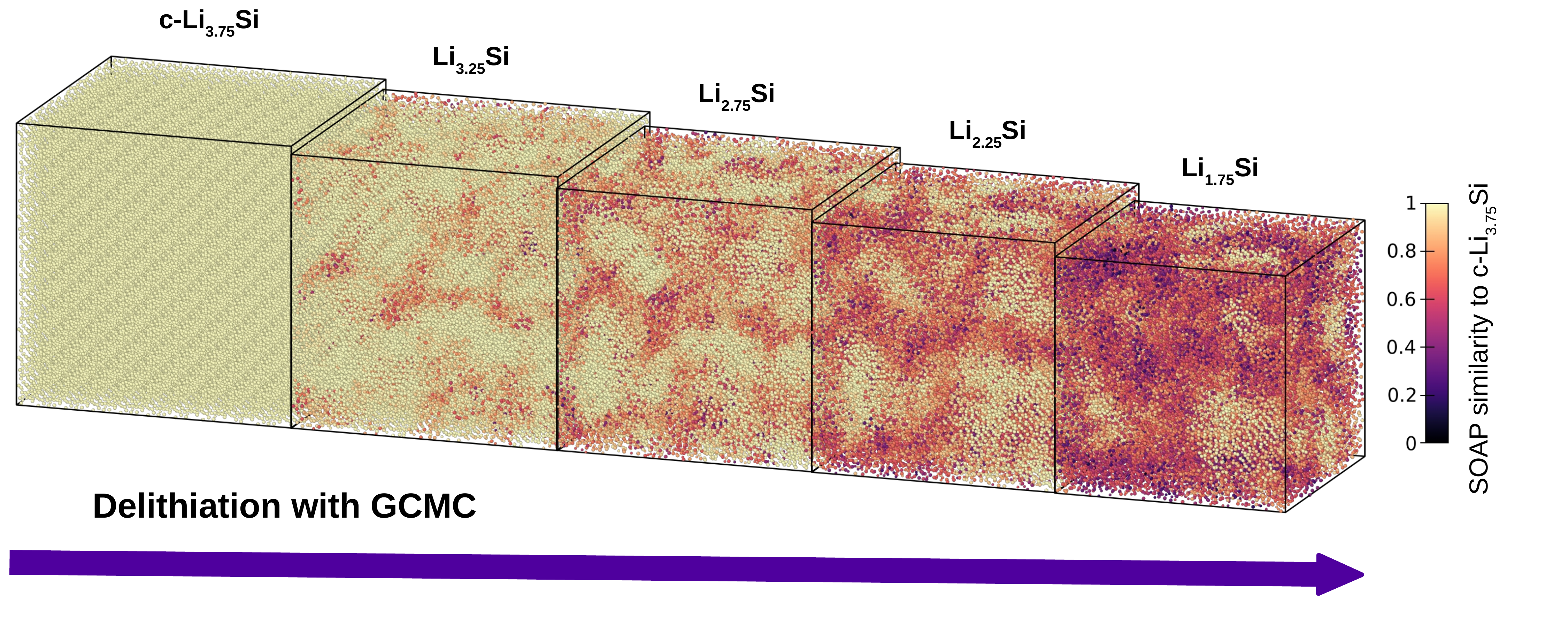}
        \label{fig:40}
    \end{subfigure}
\setlength{\unitlength}{1cm}
    \begin{picture}(0,0)
    \put(-1.0,14.8){\large\textbf{(a)}}   
    \put(6.0,14.8){\large\textbf{(b)}}
    \put(6.0,7.5){\large\textbf{(c)}}
\end{picture}
    \vspace{-0.7cm}
    \caption{(a) Stoichiometry maps showing the peak position of Si $2p^{3/2}$ for Li$_x$Si structures generated using MQA (red) and GCMC starting from c-\ce{Li_{3.75}Si} (green) and c-\ce{Li_{4.4}Si} (purple), compared to experimental data from Si electrodes obtained in \textit{operando} (black crosses) and \textit{ex situ} (blue triangles). The stars represent the values for the initial structure in the GCMC simulations and the shaded regions correspond to the predicted uncertainties. (b) Heat map showing the SOAP similarity between Si environments generated via MQA and GCMC and those in c-Li$_{3.75}$Si. The color scale represents the percentage of Si environments at a given stoichiometry that exhibit a specific SOAP similarity to c-Li$_{3.75}$Si. (c) Simulation snapshots of Li$_x$Si structures obtained during the delithiation of c-Li$_{3.75}$Si using GCMC. The color map accounts for the SOAP similarity of each local environment to c-Li$_{3.75}$Si.}
    \label{fig:stoichiometry_map}
\end{figure}

\subsubsection{Stoichiometry map for Si $2p^{3/2}$ in Li$_x$Si}
\autoref{fig:stoichiometry_map}(a) shows the stoichiometry maps for the Si $2p^{3/2}$ BE$_\text{max}$.
In the low Li concentration regime, where a-Li$_x$Si phases dominate, the Si $2p^{3/2}$ BEs resulting from the MQA (red solid line) and GCMC (green and purple solid lines) structures yield similar results, which are also comparable to experiments obtained either under \textit{operando}~\cite{endo2022} (black crosses) or \textit{ex situ}~\cite{ferraresi2016} (blue triangles) conditions. At higher Li contents, the predictions of the ML model for the GCMC-generated structures obtained from c-\ce{Li_{4.4}Si} and c-\ce{Li_{3.75}Si}, align more closely with the \textit{ex situ} measurements, where the experimentally observed downshift in BE at increasing $x$ is quantitatively well reproduced. 
It is worth noting that, as for the Li $1s$ spectrum, the experimental dataset corresponding to the black crosses was rigidly shifted to align with \textit{ex situ} data in the low-lithiation regime ($\Delta E=1.8$ eV, see Supporting Information for the details). Despite this alignment, the two experimental datasets show a discrepancy of up to $\sim$1 eV in BE$_\text{max}$ for $x>1.7$, which may arise from the different calibration approaches or from other experimental factors not accounted for in our model.~\bibnote{Ref.~\citenum{endo2022}  referenced the surface‑hydrocarbon peak in the C $1s$ region, while Ref.~\citenum{ferraresi2016} applied no energy calibration. This different calibration may result in systematic, inhomogeneous, and core‑level‑dependent energy offsets as previously observed in XPS measurements of cycled battery electrodes\cite{oswald2015}. Additional discrepancies may also originate from the finite surface depth probed by XPS, which may result in an apparent Li content that differs from the local composition experienced by the photoelectrons.}

\begin{figure}[htbp]
    \centering
        \includegraphics[width=0.5\textwidth]{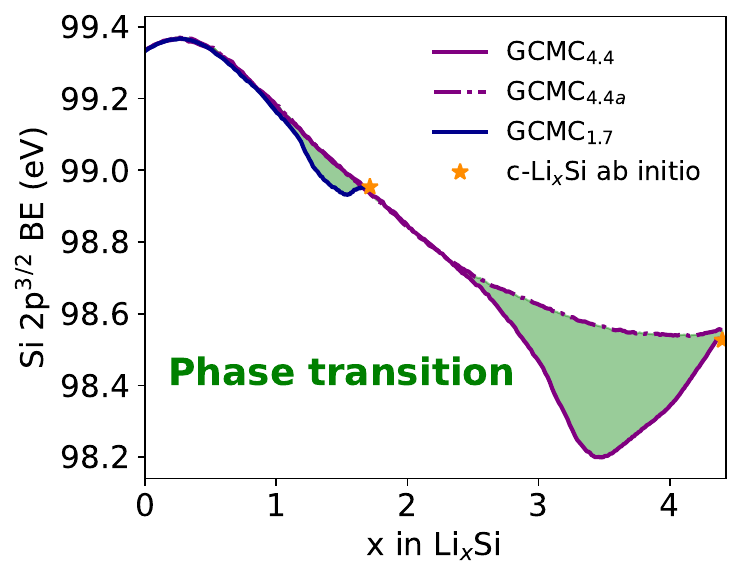}
    \caption{Stoichiometry maps for Si $2p^{3/2}$ BE$_\text{max}$ for Li$_x$Si structures generated using GCMC starting from c-\ce{Li_{4.4}Si} (purple), a-\ce{Li_{4.4}Si} (purple dashed) and c-\ce{Li_{1.7}Si} (dark blue). The values for the initial crystalline structures are represented with orange stars, while the shaded region accounts for the crystal-amorphous phase transition}
    \label{fig:Phase_transition}
\end{figure}

Interestingly, at the onset of delithiation, a pronounced downshift in the Si $2p^{3/2}$ BE$_\text{max}$ is observed for the GCMC simulations initiated from c-\ce{Li_{3.75}Si} and c-\ce{Li_{4.4}Si}, which is subsequently reversed around $x \approx 3.4$. These dips are observed exclusively when GCMC simulations use crystalline phases as initial configurations. As shown in \autoref{fig:Phase_transition}, benchmark calculations performed starting from crystalline and amorphous phases having the same Si $2p^{3/2}$ BE$_\text{max}$, namely \ce{Li_{4.4}Si} and \ce{Li_{1.7}Si}, evidence that
%
the reversal in the BE$_\text{max}$ trend is a characteristic of the trajectories initiated from the crystalline phases, indicating that this behavior originates from the crystal-to-amorphous transition. Hence, this kind of evolution observed in BE$_\text{max}$ during early delithiation can be interpreted as the spectroscopic manifestation of the structural disordering process captured by XPS (for the effects on the FWHM, see Figure S6 in Supporting Information). In particular, the Si $2p^{3/2}$ core level demonstrates greater sensitivity to this transition than the Li $1s$, highlighting its effectiveness as a probe of structural evolution. 

To support the analysis presented in the stoichiometry map, a SOAP similarity analysis was carried out, taking c-\ce{Li_{3.75}Si} as the reference system (see \autoref{fig:stoichiometry_map} (b) and (c) panels). The heat map (b) displays the percentage of Si environments that exhibit a specific
SOAP similarity to the reference system c-Li$_{3.75}$Si, comparing structures generated via MQA and GCMC. While for MQA the fraction of structures with similarity above 0.9 never exceeds 65\% at any concentration, consistent with their amorphous character, in the GCMC simulations starting from both c-Li$_{3.75}$Si and c-Li$_{4.4}$Si, the percentage stay above 90\% and 77\%, respectively, up to $x = 3.3$, indicating the presence of a large fraction of crystalline domains. This can be also seen in \autoref{fig:stoichiometry_map}(c), where few simulation snapshots are presented for the delithiation of c-Li$_{3.75}$Si using GCMC, where the color map accounts for the SOAP similarity of each local environment to c-Li$_{3.75}$Si. This analysis further reveals that delithiation proceeds in a non-uniform manner, with clusters of c-Li$_{3.75}$Si environments persisting even as the average composition (b) evolves toward amorphous states. These residual crystalline domains prevent complete Li extraction and thus provide a microscopic explanation for the irreversible capacity loss typically observed in Si-based anodes. Similar signatures of persistent structural motifs during c-Li$_{3.75}$Si delithiation have been reported previously in experimental studies \cite{boniface2016, LI2023, hernandez2025} and in MD simulations \cite{kim2017, Fu2023}. Fu et al. \cite{Fu2023}, for example, observed that the c-Li$_{3.75}$Si lattice remains largely crystalline down to c-Li$_{3.0}$Si, with limited formation of Si–Si dimers and Si–Si–Si motifs, in agreement with earlier analyses of Li-deficient c-Li$_{3.75-\delta}$Si\cite{ogata2014}. However, the simulation cell size used in that study (608 atoms) precluded the resolution of spatially heterogeneous crystalline domains. In contrast, the large-scale simulations used here (76000 atoms) enable direct visualization of these persistent crystalline clusters providing new insight into their spatial distribution, evolution and connect these structural motifs with the changes observed in the Si $2p^{3/2}$ XPS spectrum.

\section{Conclusions}

In this work, we have introduced a machine-learning framework that enables accurate, scalable, and uncertainty-aware prediction of core-level XPS binding energies in structurally complex Li$_x$Si systems, overcoming the prohibitive cost of conventional \textit{ab initio} approaches. Our model achieves near-DFT accuracy for Li $1s$ and Si $2p$ levels while remaining computationally efficient for large, amorphous configurations. The calibrated uncertainty estimates fall within typical experimental XPS resolution, providing confidence in the robustness of the predictions. Applying the model to extensive melt-quench-anneal and grand-canonical Monte Carlo simulations allowed us to construct stoichiometry maps that quantitatively link XPS peak positions to lithium content and structural evolution during (de)lithiation. Comparison with \textit{operando} and \textit{ex situ} experimental data shows excellent agreement. This analysis also reveals clear spectroscopic signatures of the crystalline-to-amorphous transition and the persistence of crystalline domains during early delithiation, offering microscopic insight into irreversible capacity loss in Si anodes. More broadly, this work establishes a general, transferable strategy for coupling machine learning, large-scale atomistic simulations, and XPS, to enable quantitative interpretation of spectroscopic data in disordered battery materials under realistic operating conditions.

\begin{acknowledgement}
This work was partly supported by the European Union’s Horizon 2020 research and innovation programme (BIG-MAP, Grant No. 957189, also part of the BATTERY 2030+ initiative, Grant No. 957213), by the European High Performance Computing Joint Undertaking (MaX Centre of Excellence -- Materials design at the eXascale, program HORIZON-EUROHPC-JU-2021-COE01, Grant No. 101093374), and by the European Union -- NextGenerationEU, through the MUR -- Prin 2022 programme (2D-FRONTIERS, Grant No. 20228879FT). 
DT and MC acknowledge support from a Sinergia grant of the Swiss National Science Foundation (grant ID CRSII5\_202296).
MC acknowledges support from the European Research Council (ERC) under the research and innovation program (Grant Agreement No. 101001890-FIAMMA) and the NCCR MARVEL, funded by the Swiss National Science Foundation (SNSF, Grant No. 205602).
The authors acknowledge the CINECA award under the ISCRA initiative, for the availability of high-performance computing resources and support. 
\end{acknowledgement}

\begin{suppinfo}
Details on: model hyperparameters and learning curves; model evaluation on the Li$_x$ DFT dataset; parameter convergence and formation energies from GCMC simulations; XPS data processing and alignment.
\end{suppinfo}

\bibliography{biblio.bib}

\newpage
\includepdf[pages=-]{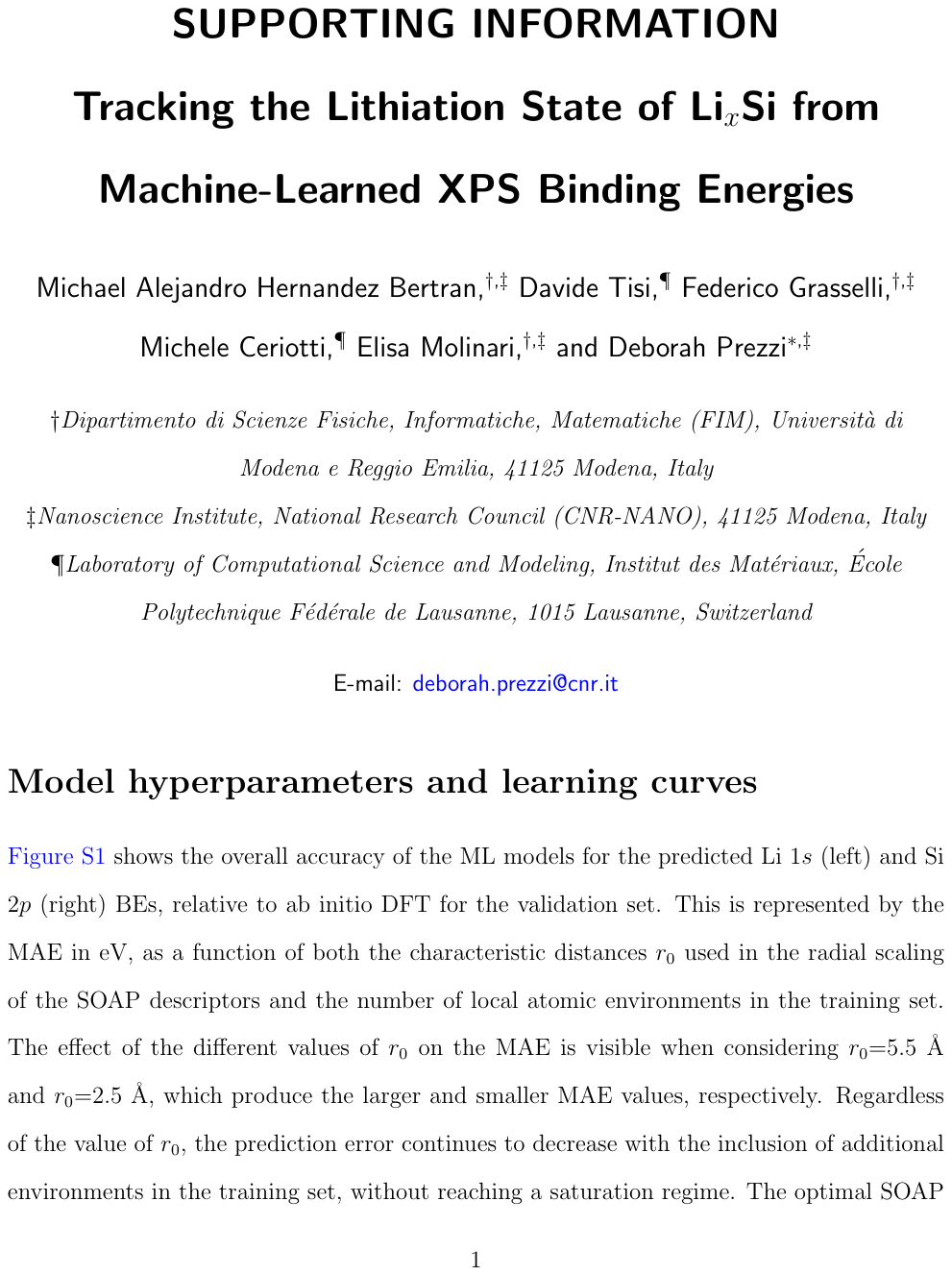}

\end{document}